\documentclass[notitlepage,nofootinbib,preprintnumbers,amssymb,superscriptaddress]{revtex4-2}
\usepackage{amsfonts,amssymb,mathtools,graphicx,color,bm}
\definecolor{ultramarine}{rgb}{0.07, 0.04, 0.56}
\definecolor{cadmiumgreen}{rgb}{0.0, 0.42, 0.24}
\definecolor{indigo(dye)}{rgb}{0.0, 0.25, 0.42}
\usepackage[linktocpage=true]{hyperref}
\hypersetup{
colorlinks=true,
citecolor=ultramarine,
linkcolor=cadmiumgreen,
urlcolor=indigo(dye),
}

\usepackage{autobreak}
\newcommand{\rinf}{r_{\rm s}}
\newcommand{\D}{{\rm d}}
\newcommand{\fr}[2]{\frac{#1}{#2}}

\newcommand{\bra}[1]{\left( #1 \right)}  
\newcommand{\brb}[1]{\left[ #1 \right]}  
  
\newcommand{\be}{\begin{equation}}  
\newcommand{\ee}{\end{equation}}
\newcommand{\bem}{\begin{bmatrix}}
\newcommand{\eem}{\end{bmatrix}}

\newcommand{\mn}{{\mu \nu}}

\begin{document}

\preprint{YITP-24-69, IPMU24-0025, RIKEN-iTHEMS-Report-24}

\title{(In)stability of the black hole greybody factors and ringdowns\\ against a small-bump correction}

\author{Naritaka Oshita}
\affiliation{Center for Gravitational Physics and Quantum Information, Yukawa Institute for Theoretical Physics, Kyoto University, 606-8502, Kyoto, Japan}
\affiliation{The Hakubi Center for Advanced Research, Kyoto University,
Yoshida Ushinomiyacho, Sakyo-ku, Kyoto 606-8501, Japan}
\affiliation{RIKEN iTHEMS, Wako, Saitama, 351-0198, Japan}

\author{Kazufumi Takahashi}
\affiliation{Center for Gravitational Physics and Quantum Information, Yukawa Institute for Theoretical Physics, Kyoto University, 606-8502, Kyoto, Japan}

\author{Shinji Mukohyama}
\affiliation{Center for Gravitational Physics and Quantum Information, Yukawa Institute for Theoretical Physics, Kyoto University, 606-8502, Kyoto, Japan}
\affiliation{Kavli Institute for the Physics and Mathematics of the Universe (WPI), The University of Tokyo Institutes for Advanced Study, The University of Tokyo, Kashiwa, Chiba 277-8583, Japan}

\begin{abstract}
Recently, it has been proposed that the black hole greybody factors can be important to model ringdown spectral amplitudes. We study the stability of greybody factors against a small-bump correction in the perturbation equation. We find (I) that the greybody factor is stable in the frequency region relevant to ringdown and (II) that it is destabilized at higher frequencies, especially for a sharper bump correction. This behavior is similar to the case of higher overtones, which is also very sensitive to a small correction. We clarify this (in)stability with the WKB analysis. As the greybody factor is stable at the frequency region relevant to the main part of ringdown, we conclude that the greybody factor is suitable to model ringdown amplitude. In order to investigate a bump correction in a self-consistent manner, we consider the small-bump correction that can be realized in the general framework of effective field theory of black hole perturbations. 
\end{abstract}

\maketitle

%%%%%%%%%%%%%%%%%%%%%%%%%%%%%%%%%%%%%%%%%%%%%%%%%%%%%%%%%%%%%%%%%%%%%%%%%%%%%%%%%%%%
%%%%%%%%%%%%%%%%%%%%%%%%%%%%%%%%%%%%%%%%%%%%%%%%%%%%%%%%%%%%%%%%%%%%%%%%%%%%%%%%%%%%
%	Introduction
%%%%%%%%%%%%%%%%%%%%%%%%%%%%%%%%%%%%%%%%%%%%%%%%%%%%%%%%%%%%%%%%%%%%%%%%%%%%%%%%%%%%
%%%%%%%%%%%%%%%%%%%%%%%%%%%%%%%%%%%%%%%%%%%%%%%%%%%%%%%%%%%%%%%%%%%%%%%%%%%%%%%%%%%%
\section{Introduction}\label{sec:intro}

The black hole ringdown is an important probe to test gravity. A ringdown waveform can be represented by a superposition of the quasinormal modes (QNMs) of a black hole. As the QNM frequencies are uniquely determined by the mass and spin of the black hole in general relativity, the black hole spectroscopy~\cite{Dreyer:2003bv,Berti:2005ys} is a promising way to perform the test of gravity in a strong field regime.
However, there are some issues in modeling ringdown with QNMs. For example, the QNM model involves many fitting parameters, as one has to assign two fitting parameters for each QNM, i.e., the amplitude and phase. This may lead to an overfitting problem~\cite{Baibhav:2023clw}, which becomes crucial when we model the ringdown waveform around the strain peak. Another issue is the instability of QNM frequencies, which is caused when we add a small correction in the wave equation governing black hole perturbations~\cite{Nollert:1996rf,Daghigh:2020jyk,Jaramillo:2020tuu,Cheung:2021bol}, though the prompt ringdown signal relevant to the time-domain observations is actually stable as shown in Refs.~\cite{Berti:2022xfj,Kyutoku:2022gbr}.

Recently, one of the authors proposed that another universal quantity, the black hole greybody factors, denoted by $\Gamma_{\ell m}$, would be useful to model ringdown amplitude~\cite{Oshita:2022pkc,Oshita:2023cjz,Okabayashi:2024qbz}, where $\ell$ and $m$ refer to angular modes. The greybody factors characterize the transmissivity through the effective potential barrier around a black hole and are uniquely determined only by the mass and spin of the black hole like QNM frequencies. With the greybody factor, one of the authors proposed that the ringdown spectral amplitude~$|\tilde{h}_{\ell m}(\omega)|$ can be modeled by
\begin{equation}
|\tilde{h}_{\ell m}(\omega)| \simeq C \sqrt{1- \Gamma_{\ell m} (\omega)}/\omega^p,
\end{equation}
where $C$ is an overall amplitude and the power~$p$ depends on the source of gravitational waves, e.g., a point particle plunging into a spinning black hole~\cite{Oshita:2023cjz} or a binary black hole leading to a spinning remnant black hole~\cite{Okabayashi:2024qbz}. 
Compared to the QNM fitting analysis, the greybody-factor model requires only a small number of fitting parameters and has been confirmed to be useful to infer the mass and spin of a remnant black hole~\cite{Oshita:2023cjz,Okabayashi:2024qbz} as the greybody factor can be extracted from the ringdown spectral amplitude for merger events 
of both extreme and comparable mass-ratio binaries.
Also, the strong decay at higher frequencies in the ringdown spectral amplitude, originating from greybody factors, can be reproduced by a destructive interference among multiple overtones~\cite{Oshita:2022pkc}. One may wonder how stable the greybody factor is if the higher overtones are destabilized by a small correction. We here consider the instability issue in the greybody-factor model by adding a small-bump (or dip) correction in the perturbation equation around a spherical black hole. 
We establish not only the stability of greybody factors in the frequency domain relevant to ringdown but also the instability of greybody factors at higher frequencies.

It has been expected that modified gravity or matter effect can be a source of the small bump in the effective potential, but no concrete model has been proposed so far.
Along this line of thought, in the present paper, we demonstrate that such a setup can be realized in the so-called effective field theory (EFT) of black hole perturbations~\cite{Mukohyama:2022enj,Mukohyama:2022skk,Khoury:2022zor,Mukohyama:2023xyf}.
The EFT was developed as an extension of the EFT of ghost condensation~\cite{Arkani-Hamed:2003pdi} and the EFT of inflation/dark energy~\cite{Cheung:2007st,Gubitosi:2012hu}, and it provides a model-independent framework to study perturbations on a black hole (or any other) background in scalar-tensor theories.
Indeed, the EFT in principle accommodates any scalar-tensor theories, e.g., Horndeski~\cite{Horndeski:1974wa,Deffayet:2011gz,Kobayashi:2011nu}, degenerate higher-order scalar-tensor (DHOST)~\cite{Langlois:2015cwa,Crisostomi:2016czh,BenAchour:2016fzp,Takahashi:2017pje,Langlois:2018jdg}, U-DHOST theories~\cite{DeFelice:2018mkq,DeFelice:2021hps,DeFelice:2022xvq} (or even more general theories proposed recently in Refs.~\cite{Takahashi:2021ttd,Takahashi:2022mew,Takahashi:2023jro,Takahashi:2023vva}), and more.
The essential assumption of the EFT is the existence of a scalar field with a timelike profile, which spontaneously breaks the time diffeomorphism invariance.
(See also Ref.~\cite{Franciolini:2018uyq} for an EFT of black hole perturbations with a spacelike scalar profile.)
The idea is that the dynamics of perturbations is described by the action that respects the residual symmetry, i.e., the spatial diffeomorphism invariance.
Once the background metric is specified, it is straightforward to expand the EFT action up to given orders in perturbations and derivatives acting on them, and the expansion coefficients tell us how the perturbations evolve.

For simplicity, we study odd-parity perturbations on a static and spherically symmetric background, for which the master equation, i.e., the generalized Regge-Wheeler (RW) equation, was derived in Ref.~\cite{Mukohyama:2022skk}.
Our input is the background metric, which determines the profile of the effective potential.
We present a concrete form of the metric that leads to a small bump (or dip) in the effective potential and study the dynamics of the perturbations based on the model.
The metric resembles the Schwarzschild metric but with a varying mass parameter:
It takes distinct constant values at the horizon and at the infinity, and these two values are smoothly interpolated by a hyperbolic tangent function.
The position, height, and width of the bump in the effective potential can be easily controlled by the parameters of the metric (see \S\ref{sec:Veff} for details).
Thus, the EFT framework allows us to study the effect of a small-bump correction in a self-consistent manner.

The rest of this paper is organized as follows.
In \S\ref{sec:setup}, 
we introduce the master equation of the perturbation, i.e., the generalized RW equation, and describe our setup.
A brief review of the EFT of black hole perturbations is given in Appendix~\ref{app:EFT_review}.
In \S\ref{sec:greybody}, we investigate the (in)stability of QNMs and greybody factors in the presence of EFT-bump correction.
Finally, we draw our conclusions in \S\ref{sec:conc}.

%%%%%%%%%%%%%%%%%%%%%%%%%%%%%%%%%%%%%%%%%%%%%%%%%%%%%%%%%%%%%%%%%%%%%%%%%%%%%%%%%%%%
%%%%%%%%%%%%%%%%%%%%%%%%%%%%%%%%%%%%%%%%%%%%%%%%%%%%%%%%%%%%%%%%%%%%%%%%%%%%%%%%%%%%
%	Setup
%%%%%%%%%%%%%%%%%%%%%%%%%%%%%%%%%%%%%%%%%%%%%%%%%%%%%%%%%%%%%%%%%%%%%%%%%%%%%%%%%%%%
%%%%%%%%%%%%%%%%%%%%%%%%%%%%%%%%%%%%%%%%%%%%%%%%%%%%%%%%%%%%%%%%%%%%%%%%%%%%%%%%%%%%
\section{Setup}\label{sec:setup}

%%%%%%%%%%%%%%%%%%%%%%%%%%%%%%%%%%%%%%%%%%
\subsection{Generalized Regge-Wheeler equation}\label{sec:RWeq}

We study odd-parity perturbations about a static and spherically symmetric background based on the EFT developed in Refs.~\cite{Mukohyama:2022enj,Mukohyama:2022skk}.
In this section, we present the generalized RW equation that governs the dynamics of the perturbations.
For a brief review of the EFT on an arbitrary background, its application to a spherically symmetric background, and the derivation of the generalized RW equation, see Appendix~\ref{app:EFT_review}.

Let us consider the background metric of the form
    \begin{align}
	\bar{g}_{\mu\nu}\D x^\mu \D x^\nu = -A(r)\D t^2 + \frac{\D r^2}{A(r)} + r^2 (\D\theta^2 + \sin^2\theta\,\D\phi^2),
	\end{align}
where $A(r)$ is a function of $r$.
Here, for simplicity, we have assumed $g_{rr}=-1/g_{tt}$.
Then, the generalized RW equation can be written as
	\begin{align}\label{eq:RW_A=B}
	\frac{\partial^2 \Psi_{\ell}}{\partial r_*^2}  - \frac{\partial^2 \Psi_{\ell}}{\partial \tilde{t}^2} - V_{{\rm eff},\ell}(r) \Psi_{\ell} = 0,
	\end{align}
where $\Psi_{\ell}$ represents the master variable for an angular mode~$\ell$, and we have defined
    \begin{align}
    r_*\coloneqq \int \frac{{\rm d}r}{F}, \qquad
    \tilde{t}\coloneqq t+ \int \frac{\sqrt{1-A}}{A}\frac{\alpha_T}{A+\alpha_T} {\rm d}r,
    \end{align}
with
    \begin{align}
    F(r)\coloneqq \frac{A+\alpha_T}{\sqrt{1+\alpha_T}}, \qquad
    \alpha_T(r)\coloneqq \frac{\alpha(r)}{M_\star^2-\alpha(r)}, \qquad
    \alpha(r)=M_\star^2+\frac{3\lambda}{r(2-2A+rA')}.
    \end{align}
Here, the parameter~$M_\star^2$ amounts to the bare Planck mass squared involved in the EFT action.
The function~$\alpha(r)$ is one of the EFT coefficients in front of tadpole terms and it has been determined by solving one of the tadpole cancellation conditions, which yields the integration constant~$\lambda$.
It should be noted that the position of the horizon associated with the odd-mode effective metric is determined by $F(r)=0$.
Note also that the function~$\alpha_T(r)$ corresponds to the relative difference between the squared sound speed of the odd mode and that of light.
The effective potential is given by
	\begin{align}\label{eq:RW_potential_A=B}
	V_{{\rm eff},\ell}(r)
    =\sqrt{1+\alpha_T}\,F\left\{\frac{\ell(\ell+1)-2}{r^2}+\frac{r}{(1+\alpha_T)^{3/4}}\left[ F\left(\frac{(1+\alpha_T)^{1/4}}{r}\right)'\,\right]'\right\},
	\end{align}
where a prime denotes the derivative with respect to $r$.
Once we fix the functional form of $A$ as well as the constant~$\lambda$, the generalized RW equation~\eqref{eq:RW_A=B} can be used to investigate the dynamics of the odd-parity perturbations.
In the following, we omit the subscript~$\ell$ unless otherwise stated.

A caveat should be mentioned here.
When one performs a time-domain analysis based on the generalized RW equation~\eqref{eq:RW_A=B} with a matter source, a constant-$\tilde{t}$ hypersurface, which is spacelike with respect to the effective metric~$-{\rm d}\tilde{t}^2+{\rm d}r_*^2$, is not necessarily a good choice of the initial surface.
Actually, the effective metric for the perturbations is different from the background metric to which matter fields minimally couple, and hence a portion of a constant-$\tilde{t}$ hypersurface can be timelike with respect to the background metric.
The radius at which a constant-$\tilde{t}$ hypersurface becomes timelike is defined by
    \begin{align}\label{eq:norm_tildet}
    \bar{g}^{\mu\nu}\partial_\mu\tilde{t}\,\partial_\nu\tilde{t}
    =-\frac{A+\alpha_T(2+\alpha_T)}{(A+\alpha_T)^2}=0.
    \end{align}
Therefore, in the presence of the matter source, one should choose another initial surface that is spacelike with respect to both the effective metric and the background metric.
(See Refs.~\cite{Nakashi:2022wdg,Nakashi:2023vul} for more detailed discussions.)

%%%%%%%%%%%%%%%%%%%%%%%%%%%%%%%%%%%%%%%%%%
\subsection{Effective potential with a small bump}\label{sec:Veff}

It is well known that a small correction in a potential barrier that appears in the wave equation in a black hole background may lead to a significantly disturbed QNM distribution~\cite{Nollert:1996rf,Daghigh:2020jyk,Jaramillo:2020tuu,Cheung:2021bol}, though the prompt ringdown signal relevant to the time-domain observations is actually stable as shown in Refs.~\cite{Berti:2022xfj,Kyutoku:2022gbr}. Such a correction has been modeled by adding a bump correction term in the wave equation by hand. On the other hand, in the EFT framework introduced in \S\ref{sec:RWeq}, such an effective potential with a bump correction can be realized by choosing, e.g.,
    \begin{align}
    A(r)=1-\fr{r_0(1+\delta)}{r}, \qquad
    \delta(r)\coloneqq \fr{\epsilon}{2}\brb{1+\tanh\bra{\fr{r-\mu}{\sigma}}},
    \end{align}
where $r_0$, $\epsilon$, $\mu$, and $\sigma$ are constants.
In what follows, we assume that $r_0$, $\mu$, and $\sigma$ are positive.
The function~$\delta(r)$ corresponds to the correction to the Misner-Sharp mass due to the nontrivial profile of the scalar field, whose magnitude is controlled by $\epsilon$.
In the limit of $r\to\infty$, the metric takes the Schwarzschild form with the horizon radius~$\rinf\coloneqq r_0(1+\epsilon)$.
Note that the horizon radius for the effective metric, which satisfies $F(r)=0$, is different from either $r_0$ or $\rinf$ in general.
In this case, the parameter~$\alpha_T$ behaves as
    \begin{align}
    \alpha_T(r)\simeq \fr{M_\star^2\rinf}{\lambda}-1,
    \end{align}
for large $r$.
Therefore, if we choose $\lambda=M_\star^2\rinf$, we have $\alpha_T\simeq0$ for large $r$ and hence the LIGO/Virgo bound~\cite{LIGOScientific:2017vwq,GBM:2017lvd,Monitor:2017mdv} can be satisfied.
(Recall that the parameter~$\alpha_{T}$ corresponds to the relative difference between the squared sound speed of gravitational waves and that of light.)
Then, the effective potential is given by
    \begin{align}
    V_{\rm eff}(r)
    &\simeq \fr{r_0(1+\delta)}{\rinf}\bra{1-\fr{\rinf}{r}}\brb{\fr{\ell(\ell+1)}{r^2}-\fr{3\rinf}{r^3}}
    -\brb{\fr{\ell(\ell+1)}{3r}+\fr{4r^2-11r_0 r+r_0^2}{6r^3}}\delta' \nonumber\\
    &\quad +\fr{(r-r_0)(5r-2r_0)}{12r^2}\delta''
    -\fr{(r-r_0)^2}{12r}\delta''',
    \label{effpot}
    \end{align}
up to the leading order in $\epsilon$.
The standard RW potential can be recovered in the limit of $\epsilon\to 0$.
For a small nonvanishing value of $\epsilon$, the effective potential has a small bump (or dip, depending on the sign of $\epsilon$) at around $r=\mu$, whose height and width are controlled by the parameters~$\epsilon$ and $\sigma$, respectively.
It should be noted that, for $\epsilon>0$, a portion of a constant-$\tilde{t}$ hypersurface can be timelike with respect to the background metric.
While this is not a problem in the present setup without matter fields, one needs to be careful in choosing proper Cauchy surfaces if matter is added to the system. 
On the other hand, for $\epsilon<0$, one can confirm that a constant-$\tilde{t}$ hypersurface is spacelike with respect to the background metric everywhere outside the horizon for the effective metric, and the above mentioned issue of Cauchy surfaces is absent with or without matter fields.

%%%%%%%%%%%%%%%%%%%%%%%%%%%%%%%%%%%%%%%%%%%%%%%%%%%%%%%%%%%%%%%%%%%%%%%%%%%%%%%%%%%%
%%%%%%%%%%%%%%%%%%%%%%%%%%%%%%%%%%%%%%%%%%%%%%%%%%%%%%%%%%%%%%%%%%%%%%%%%%%%%%%%%%%%
%	Greybody factor
%%%%%%%%%%%%%%%%%%%%%%%%%%%%%%%%%%%%%%%%%%%%%%%%%%%%%%%%%%%%%%%%%%%%%%%%%%%%%%%%%%%%
%%%%%%%%%%%%%%%%%%%%%%%%%%%%%%%%%%%%%%%%%%%%%%%%%%%%%%%%%%%%%%%%%%%%%%%%%%%%%%%%%%%%
\section{Greybody factor}\label{sec:greybody}

In this section, we study the effect of EFT corrections to the perturbation of black holes. The pseudospectrum of QNMs has been actively studied so far (see, e.g., Refs.~\cite{Nollert:1996rf,Daghigh:2020jyk,Jaramillo:2020tuu,Cheung:2021bol}) as an implication of instability of QNMs against a small correction to the angular momentum potential. We here study how another universal quantity, i.e., the greybody factor, is sensitive to the small correction by using the EFT model of black hole perturbation introduced in the previous section. One of the authors recently proposed that the greybody factor is indeed important to model the spectral amplitude of ringdown with small numbers of fitting parameters~\cite{Oshita:2022pkc,Oshita:2023cjz,Okabayashi:2024qbz}. As such, it is important to see how (in)sensitive greybody factors are against such a small correction.

\subsection{Black hole ringdown, quasinormal modes and greybody factors}
\label{sec_review_ringdown}
We numerically solve the generalized RW equation~\eqref{eq:RW_A=B} with the effective potential~\eqref{effpot} in the frequency domain. To this end, let us define the Laplace transform of $\Psi (\tilde t, r_{\ast})$ as
\begin{equation}
\psi(\omega, r_{\ast}) = \int_{\tilde{t}_0}^{\infty} {\rm d}\tilde t\, e^{i\omega \tilde{t}}\Psi(\tilde t, r_{\ast}),
\label{tr_decomposition}
\end{equation}
where $\tilde{t}_0$ is the start time of ringdown.
Then, the generalized RW equation reduces to
\begin{equation}
\left[\frac{\partial^2}{\partial r_{\ast}^2} + \omega^2 -V_{\rm eff} (r)\right] \psi(\omega, r_{\ast}) = T(\omega, r_{\ast}),
\label{g_regge_wheeler}
\end{equation}
where the function~$T(\omega,r_{\ast})$ is determined by the initial data at $\tilde t = \tilde{t}_0$:
\begin{equation}
T (\omega, r_{\ast}) \coloneqq i \omega \Psi (\tilde{t}_0, r_{\ast}) - \frac{\partial}{\partial \tilde t} \Psi (\tilde{t}_0, r_{\ast}).
\end{equation}
Using the Green's function~$G(r,r')$ to obtain a solution satisfying the following boundary condition:
\begin{align}
\psi \sim
\begin{cases}
e^{-i \omega r_{\ast}} &\text{for} \ r_{\ast} \to - \infty,\\
e^{i \omega r_{\ast}} &\text{for} \ r_{\ast} \to  \infty,
\end{cases}
\end{align}
we find that the solution is given by
\begin{equation}
\psi (\omega, r_{\ast}) = \psi_{\rm out} (\omega, r_{\ast}) \int_{- \infty}^{r_{\ast}} {\rm d}r_{\ast}' \frac{\psi_{\rm in}(\omega, r_{\ast}') T(\omega, r_{\ast}')}{2 i \omega A_{\rm in}} + \psi_{\rm in} (\omega, r_{\ast}) \int_{r_{\ast}}^{\infty} {\rm d}r_{\ast}' \frac{\psi_{\rm out}(\omega, r_{\ast}') T(\omega, r_{\ast}')}{2 i \omega A_{\rm in}},
\end{equation}
where $\psi_{\rm in}$ and $\psi_{\rm out}$ satisfy the following conditions:
\begin{align}
\psi_{\rm in} &= 
\begin{cases}
e^{-i\omega r_{\ast}} &\text{for} \ r_{\ast} \to -\infty,\\
A_{\rm out} (\omega) e^{i\omega r_{\ast}} + A_{\rm in} (\omega) e^{-i\omega r_{\ast}} &\text{for} \ r_{\ast} \to \infty,\\
\end{cases}\\
\psi_{\rm out} &= 
\begin{cases}
B_{\rm out} (\omega) e^{i\omega r_{\ast}} + B_{\rm in} (\omega) e^{-i\omega r_{\ast}} &\text{for} \ r_{\ast} \to -\infty,\\
e^{i\omega r_{\ast}} &\text{for} \ r_{\ast} \to \infty.
\end{cases}
\end{align}
In most cases, we are interested in signals at a distant region ($r_{\ast} \to \infty$), for which the solution can be approximated as
\begin{align}
\psi (\omega, r_{\ast}) &= e^{i\omega r_{\ast}} \times \frac{A_{\rm out}(\omega)}{2 i \omega A_{\rm in}(\omega)}  \times {\cal S}(\omega),
\label{spectrum_psi}
\\
{\cal S}(\omega) &= \int_{- \infty}^{\infty} {\rm d}r_{\ast}' \frac{\psi_{\rm in}(\omega, r_{\ast}') T(\omega, r_{\ast}')}{A_{\rm out}(\omega)}.
\end{align}
The greybody factor~$\Gamma(\omega)$ is defined as the transmissivity of a mode of $\omega$:
\begin{equation}
\Gamma (\omega) = \left| \frac{1}{A_{\rm in} (\omega)} \right|^2.
\label{greybody_Ain}
\end{equation}
Using the Wronskian relation of $\psi_{\rm in}$ and its complex conjugate~$\psi_{\rm in}^{\ast}$, we obtain the relation between the greybody factor~$\Gamma$ and the reflectivity~${\cal R}$:
\begin{equation}
{\cal R}(\omega) \coloneqq \left| \frac{A_{\rm out}}{A_{\rm in}} \right|^2 = 1- \Gamma (\omega).
\end{equation}
The time-domain waveform, $\Psi (\tilde{t}, r_{\ast})$, is given by the inverse Laplace transform of Eq.~(\ref{tr_decomposition}):
\begin{equation}
\Psi (\tilde t, r_{\ast}) = \int_{-\infty}^{\infty} {\rm d} \omega\, e^{-i\omega \tilde t} \psi (\omega, r_{\ast}).
\end{equation}
The function $A_{\rm out}(\omega)/ A_{\rm in}(\omega)$ has poles at $\omega = \omega_n \in {\mathbb C}$ in the half lower plane of $\omega$, which are nothing but QNMs. We then obtain the damped oscillation of $\Psi (\tilde t, r_{\ast})$ as
\begin{equation}
\Psi \sim \sum_{n} C_n e^{-i\omega_n (\tilde t -r_{\ast})},
\end{equation}
by using the residue theorem with a contour surrounding the poles and avoiding the brunch cut on the negative-imaginary axis in the complex $\omega$ plane. The excitation coefficient~$C_n$ is given by the product of ${\cal S}_n \coloneqq 
{\cal S}(\omega_n)$ and the residue of $A_{\rm out}/(2i\omega A_{\rm in})$ at $\omega = \omega_n$, i.e., the excitation factor~$E_n$:
\begin{equation}
C_n = E_n \times {\cal S}_n.
\end{equation}
The excitation factor is a universal quantity that depends only on the unique parameters of the black hole as well as the parameters of the underlying gravitational theory~\cite{Silva:2024ffz}, and is independent of the initial data and the source of perturbations. Therefore, the excitation factor~$E_n$ is regarded as an important quantity to estimate the universal {\it excitability} of each QNM.

Similarly, the spectral amplitude of ringdown~$|\psi(\omega, r_{\ast})|$ can be also decomposed into two parts, i.e., the universal part~$A_{\rm out}/(2i\omega A_{\rm in})$ and the source-dependent one~${\cal S} (\omega)$:\footnote{As shown in Eq.~\eqref{spectrum_psi}, for large $r_*$, the $r_*$-dependent part of $\psi(\omega,r_*)$ affects only the phase, and hence $|\psi(\omega, r_{\ast})|$ is a function only of $\omega$.}
\begin{equation}
|\psi(\omega, r_{\ast})| = \left| \frac{\sqrt{{\cal R}(\omega)}}{2 i \omega} \right| \times \left| {\cal S} (\omega) \right|.
\end{equation}
As ${\cal R}(\omega) = 1-\Gamma(\omega)$, the universal part in the above decomposition is determined only by the greybody factor. Recently, one of the authors has proposed that the greybody factor is another important quantity to model the spectral amplitude of ringdown and can be useful to estimate the mass and spin of a remnant black hole by fitting the greybody factor to gravitational-wave data~\cite{Oshita:2023cjz,Okabayashi:2024qbz}. In the following sections, we investigate the stability not only of QNMs but also of the greybody factor against a small bump correction that can be modeled by the framework of the EFT. The effect of an EFT correction on the QNMs and greybody factor was recently investigated in Refs.~\cite{Mukohyama:2023xyf,Konoplya:2023ppx}.
On the other hand, we focus on how a bump correction affects the two important quantities in the following.

\subsection{Instability of QNMs: pseudospectrum}
%%%%%%%%%%%%%%%%%%%%%%%%%
\begin{figure}[t]
  \centering \includegraphics[keepaspectratio=true,height=52mm]{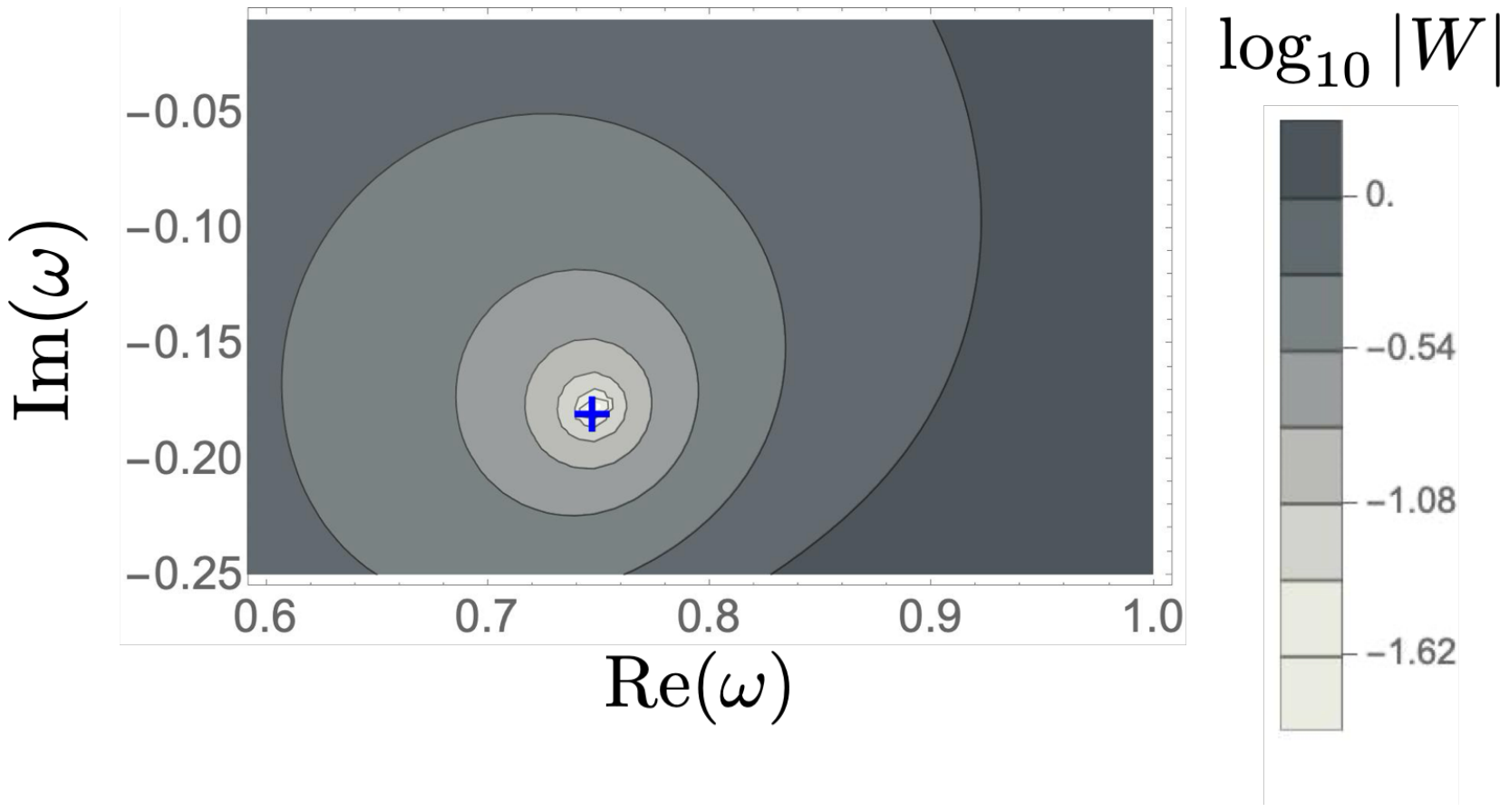}
\caption{The Wronskian~$W(\omega)$ is shown in the complex frequency plane for $\ell=2$ and $r_0=1$. As a benchmark test, we show the result for the RW equation. The frequencies at which the Wronskian~$W(\omega)$ is zero are QNM frequencies. The original fundamental QNM frequency for the quadrupole mode~$r_0 \omega_0 = 0.7473-i 0.1779$ is indicated by ``$+$'' in the plot.
}
\centering
\label{pic_funqnm}
\end{figure}
%%%%%%%%%%%%%%%%%%%%%%%%%
%%%%%%%%%%%%%%%%%%%%%%%%%
\begin{figure}[t]
  \centering \includegraphics[keepaspectratio=true,height=52mm]{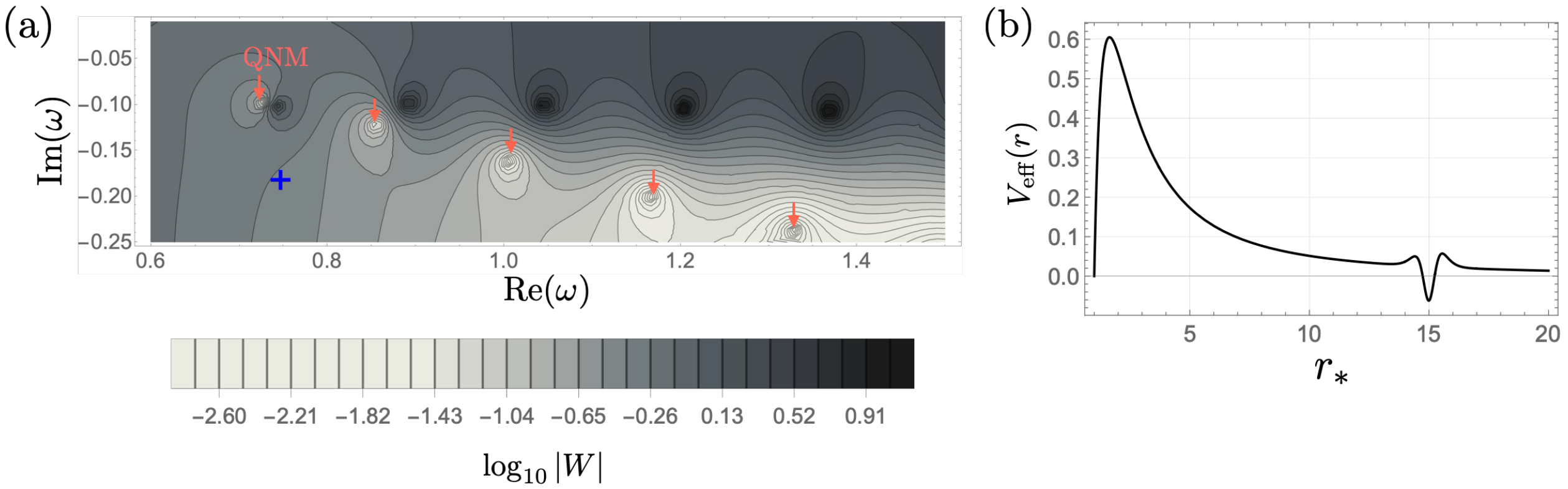}
\caption{(a) The Wronskian~$W(\omega)$ is shown in the complex frequency plane. We set $r_0=1$, $\ell=2$, $\sigma=0.5$, $\mu=15$, and $\epsilon = -0.01$. The frequencies at which the Wronskian~$W(\omega)$ is zero correspond to QNM frequencies. The original fundamental QNM frequency is indicated by ``$+$'' in the plot.
(b) The effective potential~$V(r)$ is shown for the same parameter set as in (a).
}
\centering
\label{pic_qnm}
\end{figure}
%%%%%%%%%%%%%%%%%%%%%%%%%
We here numerically investigate the instability of QNMs in our model~(\ref{effpot}) and (\ref{g_regge_wheeler}) with $\epsilon < 0$ for which the constant-$\tilde{t}$ hypersurface is timelike everywhere outside the horizon of the effective metric. We numerically solve Eq.~(\ref{g_regge_wheeler}) with $T(\omega, r_{\ast}) = 0$ to obtain the homogeneous solution. We search for specific values of $\omega$ at which the homogeneous solution satisfies the ingoing and outgoing boundary conditions at the horizon and far region, respectively. To this end, we apply the shooting method introduced by Chandrasekhar and Detweiler in Ref.~\cite{Chandrasekhar:1975zza}. We then compute the Wronskian of the ingoing and outgoing homogeneous solutions $W[\psi_{\rm in}, \psi_{\rm out}]$ over the complex plane of $\omega$, where
\begin{equation}
W[\psi_{\rm in},\psi_{\rm out}] \coloneqq \frac{{\rm d} \psi_{\rm in}}{{\rm d} r_{\ast}} \psi_{\rm out} - \frac{{\rm d} \psi_{\rm out}}{{\rm d}r_{\ast}} \psi_{\rm in}.
\end{equation}
A specific complex value of $\omega$ at which the Wronskian is zero is nothing but a QNM frequency. Our computation is precisely consistent with the value of the fundamental QNM frequency~$\omega = \omega_0$ for the quadrupole moment perturbation of the Schwarzschild black hole as is shown in Figure~\ref{pic_funqnm}. 
Turning on the small EFT correction in the potential barrier, we find that the QNM spectrum is destabilized, and many long-lived overtones appear as shown in Figure~\ref{pic_qnm}. As recognized in the literature~\cite{Nollert:1996rf,Daghigh:2020jyk,Jaramillo:2020tuu,Cheung:2021bol}, the distribution of QNMs is quite sensitive to the bump correction, which is caused by the resonance between the two potential peaks, i.e., the peak of the original potential barrier in the RW equation and the other peak of the bump correction. Due to this instability of QNMs, it is often questioned whether QNMs are really proper quantities to model ringdown waveforms. In the next section, we investigate the stability of the greybody factor against a small bump correction. The greybody factor is another important quantity to model ringdown as described in \S\ref{sec_review_ringdown}.

\subsection{Stability of greybody factors}
\begin{figure}[t]
  \centering \includegraphics[keepaspectratio=true,height=52mm]{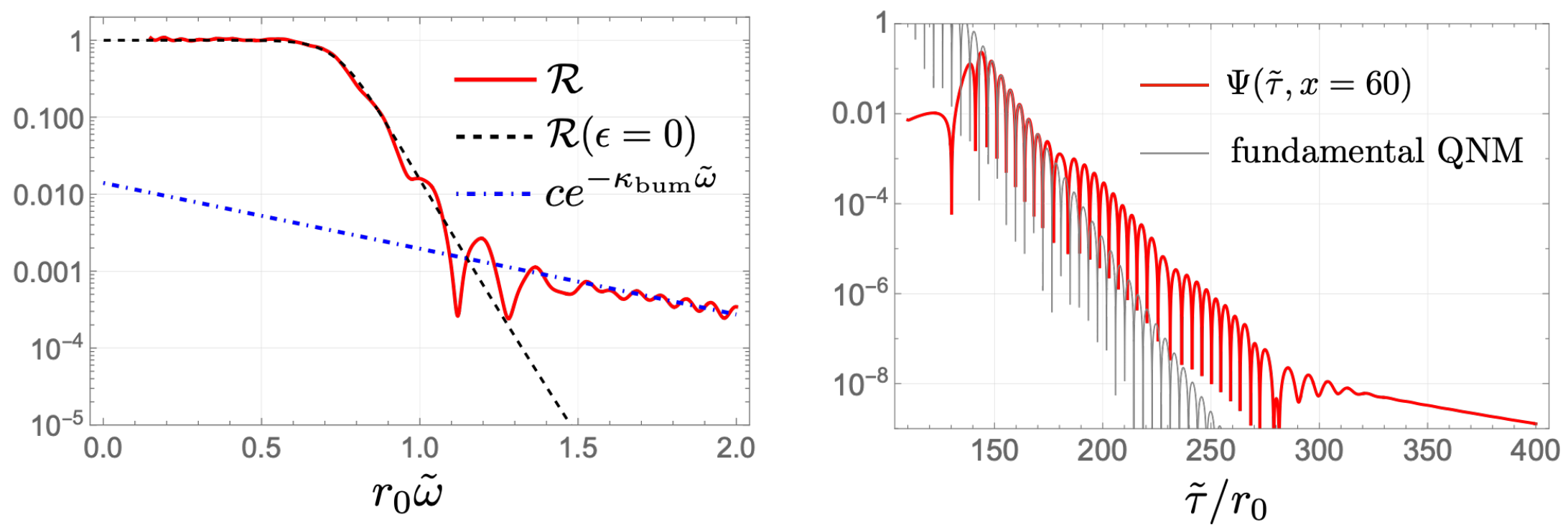}
\caption{{\bf Left}: The reflectivity~${\cal R} = 1-\Gamma$ for $\ell=2$, $\epsilon=0.01$, $\mu =15$, and $\sigma=0.5$ (red solid). The blue dot-dashed line indicates the exponential tail~(\ref{exptail}) whose exponent $\kappa_{\rm bum}$ is determined by the curvature and height of the top of the bump correction. In our parameter choice, the exponent is $\kappa_{\rm bum} \simeq 1.97$ based on Eq.~(\ref{kappa_formula}). An overall constant~$c$ is set to $0.014$. The reflectivity for $\epsilon =0$, i.e., for the Regge-Wheeler equation, is also plotted as a reference (black dashed). {\bf Right}: The time-domain waveform~$\Psi (\tau, x =60)$ (red thick solid). The thin black solid shows the damped sinusoid with the fundamental QNM frequency.
}
\centering
\label{pic_grey_eps_posi}
\end{figure}

We here compute the greybody factor with a small-bump and dip corrections in the generalized Regge-Wheeler equation. For a bump correction $\epsilon > 0$ in our EFT, the constant-$\tilde{t}$ hypersurface can be spacelike near the horizon of the effective metric. As such, imposing the proper boundary condition at the horizon is a nontrivial problem. We then solve the scattering problem by solving the partial differential equation (PDE) numerically:
\begin{equation}
{\cal A}(r) \left( {\cal F} + \frac{r_0}{r} \right) \partial_{\tilde{\tau}}^2 \Psi + \partial_x^2 \Psi - \sqrt{\frac{{\cal B}(r)^2}{4} + \frac{{\cal A}(r) r_0}{r}} \partial_{\tilde{\tau}} \partial_x \Psi + [{\cal F}(r) {\cal C}(r) - \partial_x {\cal K}(r)] \partial_x \Psi + {\cal F}(r) {\cal {\cal V}}(r) \Psi =0,
\label{eqPDE}
\end{equation}
where the explicit definition of the coordinate $\tilde{\tau}(\tilde{t},r)$, $x(r)$, and the functions ${\cal A}$, ${\cal B}$, ${\cal C}$, ${\cal K}$, ${\cal F}$, and ${\cal V}$ are given in Appendix~\ref{app:PDE_explicit_form}. We impose an initial data in which an ingoing Gaussian wave packet is initially located at a far region and eventually scattered at the angular momentum barrier ${\cal F}(r) {\cal V}(r)$ near the black hole. The initial data for $\Psi=\Psi (0,x)$ is given by
\begin{align}
\Psi (0,x) &= \cos(\Omega x) \exp\left[ -(x-x_s)^2/s^2 \right],\\
\partial_{\tilde{\tau}} \Psi (0,x) &= \left[-2 s^{-2} (x-x_s) \cos(\Omega x) - \Omega \sin(\Omega x)\right] \exp\left[ -(x-x_s)^2/s^2 \right],
\end{align}
where $\Omega$, $x_s$, and $s$ are arbitrary parameters. We set the parameters as $r_0=1$, $\Omega=1$, $x_s=80$, and $s=2$ throughout the manuscript.
We then obtain the time-domain waveform, $\Psi(\tilde{\tau},x=60)$, and compute the reflectivity ${\cal R}$ from the waveform based on the methodology used in \cite{Oshita:2021onq,Cardoso:2024qie}.
The result is shown in Figure \ref{pic_grey_eps_posi}. 
%%%%%%%%%%%%%%%%%%%%%%%%%
%%%%%%%%%%%%%%%%%%%%%%%%%
One can see that the reflectivity~${\cal R}$ for $\tilde{\omega} \lesssim \omega_0$ is insensitive to the bump correction, where $\tilde{\omega}$ is a frequency conjugate to $\tilde{\tau}$.
On the other hand, the exponential tail of the reflectivity at higher frequencies, $\tilde{\omega} \gtrsim \omega_0$, is sensitive to the small correction as it exhibits a broader exponential tail at higher frequencies (left panel in Figure~\ref{pic_grey_eps_posi}). This is consistent with the WKB analysis of the reflectivity of the bump at higher frequencies~\cite{Iyer:1986np}
\begin{align}
{\cal R} \sim e^{-\kappa_{\rm bum} \tilde{\omega} },
\label{exptail}
\end{align}
where, at the leading order, the exponent factor is given by
\begin{equation}
\kappa_{\rm bum} = \frac{2 \pi \sqrt{2V_{\rm eff}(r_{*{\rm bum}})}}{\sqrt{-\,\partial_{r_*}^2V_{\rm eff}(r_{*{\rm bum}})}}.
\label{kappa_formula}
\end{equation}
Even if the height of the potential is small, it may destabilize the greybody factor at higher frequencies if the curvature at the top of the bump is larger than or comparable to that of the potential barrier as it leads to the smaller value of $\kappa_{\rm bum}$ and the tail becomes broader.
Therefore, as a result of the correction, the reflectivity~${\cal R}$ may have a broader exponential tail at higher frequencies with ${\cal R} \sim e^{-\kappa_{\rm bum} \tilde{\omega}}$.
Also, the oscillation in the reflectivity~${\cal R} (\tilde{\omega})$ is caused by the resonance in the {\it cavity} between the potential barrier and the small bump. For a bump that makes a cavity with the length of $\mu$, the resonance appears with the resonant frequency of $\sim \pi/\mu$. 
The resonance can be seen in the time-domain data as well (right panel of Figure~\ref{pic_grey_eps_posi}).
In both the greybody factors and the time-domain waveform, the effect of the small correction appears at the same order~$\sim 10^{-3}$. It implies that the level of the stability of greybody factor would be comparable with that of the time-domain waveform.
%%%%%%%%%%%%%%%%%%%%%%%%%
\begin{figure}[t]
  \centering \includegraphics[keepaspectratio=true,height=52mm]{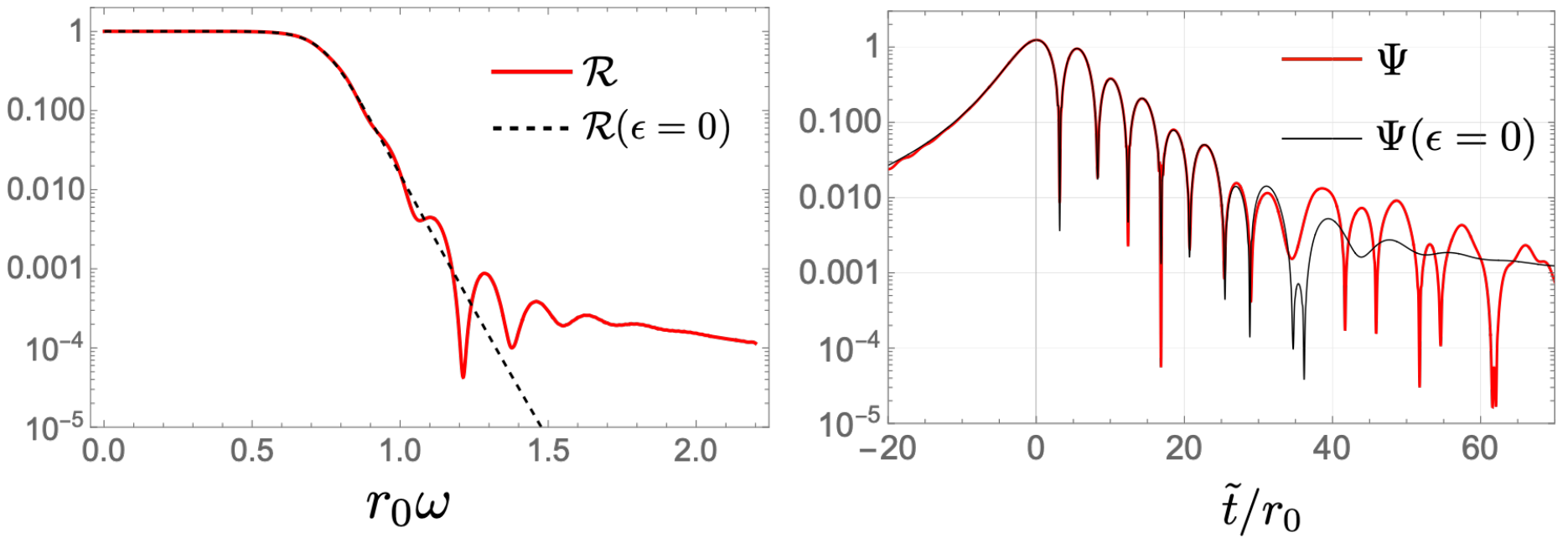}
\caption{{\bf Left}: The reflectivity~${\cal R}$ (red solid) numerically obtained for $\ell=2$, $\epsilon=-0.01$, $\mu = 15$, and $\sigma=0.5$ which is the same as the parameters in Figure \ref{pic_qnm}. {\bf Right}: The time-domain waveform for the same parameters in the left panel. The time-domain data for the Regge-Wheeler equation ($\epsilon=0$) is also plotted as a reference.
}
\centering
\label{pic_stability}
\end{figure}
%%%%%%%%%%%%%%%%%%%%%%%%%

Let us see the stability of the greybody factor in the same situation shown in Figure \ref{pic_qnm}, where the QNM distribution is significantly disturbed by the small correction with the negative small value of $\epsilon$. To compute the greybody factor for a dip correction~$\epsilon < 0$, we numerically obtain the homogeneous solution to the differential equation~(\ref{g_regge_wheeler}) by imposing the ingoing boundary condition near the horizon. This is justified as the constant-$\tilde{t}$ surface is timelike everywhere outside the horizon. We then read off the asymptotic amplitudes~$A_{\rm in}$ and $A_{\rm out}$ by using the Wronskian and obtain the greybody factor~(\ref{greybody_Ain}). Figure~\ref{pic_stability} shows the reflectivity, ${\cal R} = 1-\Gamma$, with respect to frequency~$\omega$. The stability and instability of the greybody factor is found to be similar to the case of $\epsilon > 0$. 
Also, We obtain the time-domain data by performing the inverse Laplace transform of (\ref{spectrum_psi}) with ${\cal S} (\omega) =1$. This corresponds to the computation of ringdown in which each QNM is excited with the amplitude of the excitation factor.
We then find that the effect of the small correction appears at the same order in both the greybody factor and the time-domain waveform $\lesssim 10^{-2}$. 
Nevertheless, the QNM distribution is significantly disturbed and the original fundamental mode is absent as is shown in Figure \ref{pic_qnm} with the same parameter choice.

%%%%%%%%%%%%%%%%%%%%%%%%%%%%%%%%%%%%%%%%%%%%%%%%%%%%%%%%%%%%%%%%%%%%%%%%%%%%%%%%%%%%
%%%%%%%%%%%%%%%%%%%%%%%%%%%%%%%%%%%%%%%%%%%%%%%%%%%%%%%%%%%%%%%%%%%%%%%%%%%%%%%%%%%%
%	Conclusions
%%%%%%%%%%%%%%%%%%%%%%%%%%%%%%%%%%%%%%%%%%%%%%%%%%%%%%%%%%%%%%%%%%%%%%%%%%%%%%%%%%%%
%%%%%%%%%%%%%%%%%%%%%%%%%%%%%%%%%%%%%%%%%%%%%%%%%%%%%%%%%%%%%%%%%%%%%%%%%%%%%%%%%%%%
\section{Discussions and Conclusions}\label{sec:conc}
We have investigated the stability of the greybody factors against a small bump correction. We have found (I)~that the greybody factor is stable in the frequency region relevant to ringdown and (II)~that it is destabilized at higher frequencies, especially for a sharper bump correction (see Figures~\ref{pic_grey_eps_posi} and \ref{pic_stability}). We have confirmed that the destabilized part, i.e., an exponential tail of the greybody factor, is consistent with the WKB analysis.
However, we have confirmed that such a destabilization occurs at higher frequencies. We then conclude that the greybody factors would be useful to model ringdown waveform as was proposed by one of the authors in Refs.~\cite{Oshita:2022pkc,Oshita:2023cjz,Okabayashi:2024qbz}. 
For the QNM spectrum, higher overtones are very unstable against a small bump correction. We would argue that it may be relevant to the destabilization of the greybody factors at higher frequencies that we have discovered in this paper. The exponential decay of the ringdown spectral amplitude at high frequencies, originating from greybody factors, can be reconstructed by superposed multiple higher overtones as was demonstrated in Ref.~\cite{Oshita:2022pkc}. As such, destabilized overtones may cause the destabilization of the greybody factor at higher frequencies.
In this sense, if overtones are indeed destabilized, it might be possible to prove it by observing ringdown spectral amplitude at higher frequencies and it may exhibit a broader exponential decay in the spectral amplitude, though it would need a high-precision observation of gravitational waves.

Moreover, we have formulated how a small bump correction could appear in the framework of effective field theory of black hole perturbations with a timelike profile developed in Refs.~\cite{Mukohyama:2022enj,Mukohyama:2022skk}. 
We have confirmed the (in)stability of the greybody factors and QNMs in the framework leading to a small bump correction in a self-consistent manner.
Our results would hold qualitatively for any situation where the wave equation is subjected to a small-bump correction.

\vskip3mm
{\bf Note added}: While we were finalizing this project, we noticed the paper~\cite{Rosato:2024arw} by Romeo Felice Rosato, Kyriakos Destounis, and Paolo Pani on a similar subject.
Their results are qualitatively consistent with ours.

%%%%%%%%%%%%%%%%%%%%%%%%%%%%%%%%%%%%%%%%%%%%%%%%%%%%%%%%%%%%%%%%%%%%%%%%%%%%%%%%%%%%
%%%%%%%%%%%%%%%%%%%%%%%%%%%%%%%%%%%%%%%%%%%%%%%%%%%%%%%%%%%%%%%%%%%%%%%%%%%%%%%%%%%%

\acknowledgments{
N.O.~was supported by Japan Society for the Promotion of Science (JSPS) KAKENHI Grant No.~JP23K13111 and by the Hakubi project at Kyoto University.
K.T.~was supported by JSPS KAKENHI Grant No.~JP23K13101.
S.M.~was supported in part by JSPS Grants-in-Aid for Scientific Research No.~24K07017 and the World Premier International Research Center Initiative (WPI), MEXT, Japan. 
}

%%%%%%%%%%%%%%%%%%%%%%%%%%%%%%%%%%%%%%%%%%%%%%%%%%%%%%%%%%%%%%%%%%%%%%%%%%%%%%%%%%%%
%%%%%%%%%%%%%%%%%%%%%%%%%%%%%%%%%%%%%%%%%%%%%%%%%%%%%%%%%%%%%%%%%%%%%%%%%%%%%%%%%%%%
\appendix
\section{Review of the EFT}\label{app:EFT_review}

\subsection{EFT action}

Let us briefly review the EFT of perturbations on an arbitrary background with a timelike scalar profile developed in Refs.~\cite{Mukohyama:2022enj,Mukohyama:2022skk,Mukohyama:2023xyf}.
The main idea of the EFT is that the scalar field~$\Phi$ has a timelike profile at the background, which spontaneously breaks the time diffeomorphism invariance and defines a preferred slicing of spacetime given by $\Phi={\rm const}$, similarly to the EFT of ghost condensation~\cite{Arkani-Hamed:2003pdi} and the EFT of inflation/dark energy~\cite{Cheung:2007st,Gubitosi:2012hu}.
The unit normal vector associated with a constant-$\Phi$ hypersurface is defined as
    \begin{align}\label{eq:normal_EFT}
    n_\mu \coloneqq
    -\frac{\partial_\mu \Phi}{\sqrt{-X}}
    \rightarrow - \frac{\delta_\mu^\tau}{\sqrt{-g^{\tau\tau}}},
    \end{align}
with $X\coloneqq g^{\mu\nu}\partial_\mu\Phi\partial_\nu\Phi$ being the kinetic term of the scalar field, so that $n_\mu n^\mu = -1$.
Here, the expression to the right of the arrow is the one in the unitary gauge where $\Phi$ is spatially uniform, and $\tau$ is the time coordinate.
In the unitary gauge, the residual symmetry of the EFT is only the spatial diffeomorphism invariance, and therefore the EFT action is written in terms of not only objects with spacetime covariance but also spatially covariant objects.
For instance, the time coordinate~$\tau$ and $g^{\tau\tau}$ behave as scalars under spatial diffeomorphisms, and hence are included in the building blocks of the EFT.
Also, the extrinsic curvature defined by
    \be
    K_\mn\coloneqq h_\mu^\rho \nabla_\rho n_\nu,
    \ee
with $h_\mn\coloneqq g_\mn+n_\mu n_\nu$ being the induced metric on a constant-$\tau$ hypersurface, is also one of the building blocks.
On top of these, either of the spacetime Riemann tensor~$\tilde{R}_{\mu\nu\alpha\beta}$ and the spatial Riemann tensor~${}^{(3)}\!R_{\mu\nu\alpha\beta}$ can be included in the EFT action.\footnote{We put a tilde on the spacetime curvature tensor so that the notation matches with that in \cite{Mukohyama:2022skk,Mukohyama:2022enj}. Similarly, the spacetime curvature scalar will be denoted by $\tilde{R}$.}
It should be noted that $\tilde{R}_{\mu\nu\alpha\beta}$ and ${}^{(3)}\!R_{\mu\nu\alpha\beta}$ are related to each other via the Gauss-Codazzi-Ricci equations, and we choose the latter as an independent building block.
Note also that one can always rewrite ${}^{(3)}\!R_{\mu\nu\alpha\beta}$ in terms of the spatial Ricci tensor~${}^{(3)}\!R_{\mu\nu}$ thanks to the fact that the Weyl tensor vanishes in three dimensions.
Therefore, the action under the unitary gauge can be written in the form 
    \begin{align}
    S = \int \D^4x \sqrt{-g}\,F(g^{\tau\tau}, K^\mu_\nu, {}^{(3)}\!R^\mu_\nu, \nabla_\mu, \tau),
    \label{eq:action_uni}
    \end{align}
where $F$ is a scalar function constructed out of its arguments.
Note that we have not specified the background metric, and hence the action~\eqref{eq:action_uni} can be applied to arbitrary background geometries.
Having said that, in practice, we impose some symmetries on the background (e.g., spherical symmetry) and expand the action~\eqref{eq:action_uni} up to the necessary order in perturbations and derivatives.

Let us now define the perturbations about an inhomogeneous background as follows:
    \begin{align}
    \delta g^{\tau\tau} \coloneqq g^{\tau\tau} - \bar{g}^{\tau\tau}(\tau, x^i), \qquad
    \delta K^\mu_\nu \coloneqq K^\mu_\nu - \bar{K}^\mu_\nu(\tau, x^i), \qquad
    \delta {}^{(3)}\!R^\mu_\nu \coloneqq {}^{(3)}\!R^\mu_\nu -  {}^{(3)}\!\bar{R}^\mu_\nu(\tau, x^i),
    \label{eq:pert}
    \end{align}
where a bar denotes the background value.
The EFT action is written as a polynomial of these perturbation variables as well as their derivatives.
We note that each EFT coefficient can depend on $x^i$ due to the inhomogeneity of the background, which apparently breaks the spatial diffeomorphism invariance of the EFT action.
However, the EFT action as a whole should be invariant under spatial diffeomorphisms since we are just Taylor-expanding the action~\eqref{eq:action_uni} that is manifestly spatial diffeomorphism invariant.
This suggests that the EFT coefficients are related to each other via some particular relations, i.e., the consistency relations~\cite{Mukohyama:2022enj}, to maintain the spatial diffeomorphism invariance.
The consistency relations can be obtained by applying the chain rule to the spatial derivative of the Taylor coefficients.
For example, one obtains
    \begin{align}
    \frac{\partial}{\partial x^i}\bar{F}
    &=\left.\fr{\D}{\D x^i}F(g^{\tau\tau}, K^\mu_\nu, {}^{(3)}\!R^\mu_\nu, \nabla_\mu, \tau)\right|_{\rm BG} \nonumber \\
    &= \bar{F}_{g^{\tau\tau}} \frac{\partial \bar{g}^{\tau\tau}}{\partial x^i} + \bar{F}_{K^\mu_\nu} \frac{\partial \bar{K}^\mu_\nu}{\partial x^i} + \bar{F}_{{}^{(3)}\!R^\mu_\nu} \frac{\partial {}^{(3)}\!\bar{R}^\mu_\nu}{\partial x^i}+\cdots,
    \end{align}
where the notation~$\bar{F}_Q \coloneqq \partial F/\partial Q|_{\rm BG}$ refers to the Taylor coefficient evaluated on the background and we have omitted terms of higher order in derivatives.
This equation puts a constraint on the Taylor coefficients~$\bar{F}$, $\bar{F}_{g^{\tau\tau}}$, $\bar{F}_{K^\mu_\nu}$, and $\bar{F}_{{}^{(3)}\!R^\mu_\nu}$ (as well as those associated with higher-derivative terms if they exist).
In general, one obtains infinitely many such consistency relations by applying the chain rule to the spatial derivative of other Taylor coefficients.
Note that the chain rule with respect to the $\tau$-derivative does not lead to a nontrivial constraint because the action depends explicitly on $\tau$ in general.
However, if we restrict ourselves to shift-symmetric scalar-tensor theories, then the $\tau$-dependence of the action is prohibited and the chain rule with respect to the $\tau$-derivative results in an additional set of infinitely many consistency relations on the EFT coefficients~\cite{Khoury:2022zor} (see also Ref.~\cite{Finelli:2018upr} for a similar constraint in the case of shift-symmetric EFT of inflation).\footnote{By gauging the shift symmetry~\cite{Cheng:2006us}, one obtains EFT of vector-tensor theories~\cite{Aoki:2021wew,Aoki:2023bmz}. In this case, the action is invariant under a combined $U(1)$ and time diffeomorphism, and therefore one has to impose a set of consistency relations associated with it.}

Now we are ready to write down the EFT action. For simplicity, following Ref.~\cite{Mukohyama:2022skk}, we consider a minimal EFT action that accommodates the shift-symmetric quadratic higher-order scalar-tensor theories.
Then, the EFT action can be written in the form\footnote{We put a tilde on $\beta$ and $M_1$ so that the notation matches with that in \cite{Mukohyama:2022enj}.}
    \begin{align}\label{eq:EFT_shift_Z2}
	S = \int \D^4x \sqrt{-g} \bigg[&\frac{M_\star^2}{2}
	R - \Lambda(y) - c(y)g^{\tau\tau} - \tilde{\beta}(y) K - \alpha(y)\bar{K}^\mu_\nu K^{\nu}_\mu -\zeta(y) n^\mu\partial_\mu g^{\tau\tau} \nonumber \\ 
	& + \frac{1}{2} m_2^4(y) (\delta g^{\tau\tau})^2 + \frac{1}{2} \tilde{M}_1^3(y) \delta g^{\tau\tau} \delta K + \frac{1}{2} M_2^2(y) \delta K^2 + \frac{1}{2} M_3^2(y) \delta K^\mu_\nu \delta K^\nu_\mu \nonumber \\
	& + \frac{1}{2}\mu_1^2(y) \delta g^{\tau\tau} \delta {}^{(3)}\!R + \frac{1}{2} \lambda_1(y)^\mu_\nu \delta g^{\tau\tau} \delta K^\nu_\mu + \frac{1}{2}{\mathcal M}_1^2(y)(\bar{n}^\mu\partial_\mu\delta g^{\tau\tau})^2 \nonumber \\
	& +\frac{1}{2}{\mathcal M}_2^2(y)\delta K(\bar{n}^\mu\partial_\mu\delta g^{\tau\tau})+\frac{1}{2}{\mathcal M}_3^2(y)\bar{h}^{\mu\nu}\partial_\mu\delta g^{\tau\tau}\partial_\nu\delta g^{\tau\tau}+\cdots \bigg],
	\end{align}
where $y = (\tau, x^i)$ and we have defined
    \begin{align}\label{eq:4d_Ricci}
    R \coloneqq {}^{(3)}\!R + K_{\mu\nu}K^{\mu\nu} - K^2 = \tilde{R} - 2\nabla_\mu( K n^\mu - n^\nu\nabla_\nu n^\mu),
    \end{align}
with $\tilde{R}$ being the spacetime Ricci scalar.
Here, we work in a frame where the coefficient in front of $R$ is a constant, which we denote by $M_\star^2/2$.
The explicit expression of the consistency relations mentioned above as well as the dictionary with concrete theories (e.g., Horndeski theories) can be found in Refs.~\cite{Mukohyama:2022skk,Mukohyama:2022enj}.
To reiterate, so far we have not assumed any specific forms of $\bar{\Phi}(\tau)$ and $\bar{g}_{\mu\nu}$, and hence the action~\eqref{eq:EFT_shift_Z2} applies to an arbitrary background with a timelike scalar profile.

%%%%%%%%%%%%%%%%%%%%%%%%%%%%%%%%%%%%%%%%%%
\subsection{Static and spherically symmetric background}

Let us consider the following static and spherically symmetric background metric:
    \begin{align}\label{eq:metric_BG}
	\bar{g}_{\mu\nu}\D x^\mu \D x^\nu = -A(r)\D t^2 + \frac{\D r^2}{B(r)} + r^2 (\D\theta^2 + \sin^2\theta\,\D\phi^2),
	\end{align}
where $A(r)$ and $B(r)$ are functions of $r$.
The metric can be equivalently written in terms of the so-called Lema\^{\i}tre coordinates~\cite{Lemaitre:1933gd,Khoury:2020aya} as
	\begin{align}\label{eq:BG_Lemaitre}
	\bar{g}_{\mu\nu}\D x^\mu \D x^\nu = -\D\tau^2 + [1 - A(r)] \D\rho^2 + r^2 (\D\theta^2 + \sin^2\theta\,\D\phi^2),
	\end{align}
where the coordinates~$\tau$ and $\rho$ are respectively related to $t$ and $r$ via 
	\begin{align}
	\D\tau = \D t + \sqrt{\frac{1 - A}{AB}}\,\D r, \qquad
    \D\rho = \D t + \frac{\D r}{\sqrt{AB(1 - A)}}. \label{eq:trans_Lemai}
	\end{align}
Note that $r$ depends on $\tau$ and $\rho$ only through the combination~$\rho - \tau$ and
    \begin{align}
    \partial_\rho r = - \dot{r} = \sqrt{\frac{B(1 - A)}{A}},
    \end{align}
where a dot denotes the derivative with respect to $\tau$.

We now assume that the EFT respects the shift symmetry of the scalar field, which prohibits the explicit $\tau$-dependence in Eq.~\eqref{eq:action_uni}.
Then, thanks to the static and spherically symmetric background, the EFT coefficients in the action~\eqref{eq:EFT_shift_Z2} are functions of $r$ only.
Written explicitly, we consider the following EFT action:
    \begin{align}
	S = \int \D^4x \sqrt{-g} \bigg[&\frac{M_\star^2}{2}R - \Lambda(r) - c(r)g^{\tau\tau} - \tilde{\beta}(r) K - \alpha(r)\bar{K}^\mu_\nu K^{\nu}_\mu -\zeta(r) \bar{n}^\mu\partial_\mu g^{\tau\tau} \nonumber \\ 
	& + \frac{1}{2} m_2^4(r) (\delta g^{\tau\tau})^2 + \frac{1}{2} \tilde{M}_1^3(r) \delta g^{\tau\tau} \delta K + \frac{1}{2} M_2^2(r) \delta K^2 + \frac{1}{2} M_3^2(r) \delta K^\mu_\nu \delta K^\nu_\mu \nonumber \\
	& + \frac{1}{2}\mu_1^2(r) \delta g^{\tau\tau} \delta {}^{(3)}\!R + \frac{1}{2} \lambda_1(r)^\mu_\nu \delta g^{\tau\tau} \delta K^\nu_\mu + \frac{1}{2}{\mathcal M}_1^2(r)(\bar{n}^\mu\partial_\mu\delta g^{\tau\tau})^2 \nonumber \\
	& +\frac{1}{2}{\mathcal M}_2^2(r)\delta K(\bar{n}^\mu\partial_\mu\delta g^{\tau\tau})+\frac{1}{2}{\mathcal M}_3^2(r)\bar{h}^{\mu\nu}\partial_\mu\delta g^{\tau\tau}\partial_\nu\delta g^{\tau\tau}+\cdots \bigg].
	\label{eq:EFT_HOST}
	\end{align}
Note that only the terms in the first line are relevant for the tadpole cancellation conditions (or the background equations of motion) for the metric.
The nontrivial components read
    \begin{equation}
	\begin{split}
	\Lambda-c
	&=M_\star^2(\bar{G}^\tau{}_\rho-\bar{G}^\rho{}_\rho), \\
	\Lambda+c+\frac{2}{r^2}\sqrt{\frac{B}{A}}\left(r^2\sqrt{1-A}\,\zeta\right)'
	&=-M_\star^2\bar{G}^\tau{}_\tau, \\
	\left[\partial_\rho\bar{K}+\frac{1-A}{r}\left(\frac{B}{A}\right)'\,\right]\alpha+\frac{A'B}{2A}\alpha'+\sqrt{\frac{B(1 - A)}{A}}\tilde{\beta}'
	&=-M_\star^2\bar{G}^\tau{}_\rho, \\
	\frac{1}{2r^2}\sqrt{\frac{B}{A}}\left[r^4\sqrt{\frac{B}{A}}\left(\frac{1-A}{r^2}\right)'\alpha\right]'
	&=M_\star^2(\bar{G}^\rho{}_\rho-\bar{G}^\theta{}_\theta),
	\end{split} \label{EOM_BG}
    \end{equation}
with the relevant components of the background Einstein tensor given by
    \begin{equation}
	\begin{split}
	&\bar{G}^\tau{}_\tau=-\frac{[r(1-B)]'}{r^2}+\frac{1-A}{r}\left(\frac{B}{A}\right)', \qquad
	\bar{G}^\tau{}_\rho=-\frac{1-A}{r}\left(\frac{B}{A}\right)', \\
	&\bar{G}^\rho{}_\rho=-\frac{[r(1-B)]'}{r^2}-\frac{1}{r}\left(\frac{B}{A}\right)', \qquad
	\bar{G}^\theta{}_\theta=\frac{B(r^2A')'}{2r^2A}+\frac{(r^2A)'}{4r^2}\left(\frac{B}{A}\right)',
	\end{split}
    \end{equation}
where a prime denotes the derivative with respect to $r$.
Note that, when $A(r)=B(r)$, the fourth equation can be integrated to yield
    \begin{equation}\label{eq:alpha2Mstar}
    \alpha=M_\star^2+\frac{3\lambda}{r(2-2A+rA')},
    \end{equation}
with $\lambda$ being a constant.
The above equations provide relations among the EFT coefficients through given functions~$A(r)$ and $B(r)$.

%%%%%%%%%%%%%%%%%%%%%%%%%%%%%%%%%%%%%%%%%%
\subsection{Odd-parity perturbations}

Let us first define the perturbation variables associated with the odd-parity (or axial) perturbations.
Thanks to the spherical symmetry of the background metric~\eqref{eq:metric_BG}, it is convenient to decompose the perturbations in terms of the spherical harmonics.
The odd-parity perturbations~$\delta g_{\mu\nu} \coloneqq g_{\mu\nu} - \bar{g}_{\mu\nu}$ can be expressed as follows:
    \begin{align}
    \begin{split}
    \delta g_{\tau\tau} &= \delta g_{\tau\rho} = \delta g_{\rho\rho} = 0 , \\
    \delta g_{\tau a} &= \sum_{\ell,m} r^2 h_{0,\ell m}(\tau,\rho) E_a^{\ b} \bar{\nabla}_b Y_{\ell m}(\theta, \phi) , \\
    \delta g_{\rho a} &= \sum_{\ell,m} r^2 h_{1,\ell m}(\tau,\rho) E_a^{\ b} \bar{\nabla}_b Y_{\ell m}(\theta, \phi) , \\
    \delta g_{ab} &= \sum_{\ell,m} r^2 h_{2,\ell m}(\tau,\rho) E_{(a|}^{\ \ \ c} \bar{\nabla}_c \bar{\nabla}_{|b)} Y_{\ell m}(\theta, \phi),
    \end{split}
    \end{align}
where $Y_{\ell m}$ is the spherical harmonics, $E_{ab}$ is the completely anti-symmetric rank-2 tensor defined on a $2$-sphere, $\bar{\nabla}_a$ is the covariant derivative with respect to the unit $2$-sphere metric, and $a, b, \cdots\in\{\theta, \phi\}$.
We note that the odd-parity perturbations are nontrivial for $\ell\ge 1$, and the function~$h_2$ is absent for $\ell=1$.

Let us now consider the EFT action which is relevant for the odd-parity perturbations.
Since parity-even operators like $\delta g^{\tau\tau}$ and $\delta K$ are at least quadratic in the odd-parity perturbations, the EFT action for the odd sector is written as
	\begin{align}
	S_{\rm odd} = \int {\rm d}^4x \sqrt{-g} \bigg[\frac{M_\star^2}{2}R - \Lambda(r) - c(r)g^{\tau\tau} -\tilde{\beta}(r) K - \alpha(r)\bar{K}^{\mu}_\nu K^\nu_{\mu} -\zeta(r) \bar{n}^\mu\partial_\mu g^{\tau\tau} + \frac{1}{2} M_3^2(r) \delta K^\mu_\nu \delta K^\nu_\mu \bigg], \label{eq:EFT_action}
	\end{align}
up to the second order in the perturbations.
We shall use this action to derive the quadratic action for the odd-parity perturbations.
In doing so, one usually performs a gauge fixing to remove a redundant degree of freedom.
Let us consider an infinitesimal coordinate transformation~$x^\mu \rightarrow x^\mu + \epsilon^\mu$ with
    \begin{align}
    \epsilon^\tau = \epsilon^\rho = 0, \qquad \epsilon^a = \sum_{\ell,m} \Xi_{\ell m}(\tau,\rho) E^{ab} \bar{\nabla}_b Y_{\ell m}(\theta,\phi),
    \end{align}
the perturbations~$h_0$, $h_1$, and $h_2$ are transformed as
    \begin{align}
    h_0 \rightarrow h_0 - \dot{\Xi}, \qquad h_1 \rightarrow h_1 - \partial_\rho \Xi, \qquad h_2 \rightarrow h_2 - 2 \Xi.
    \end{align}
Namely, the function~$\Xi$ corresponds to the gauge degree of freedom in the odd-parity sector.
We are mostly interested in modes with $\ell\ge 2$, where we expect to have one dynamical degree of freedom corresponding to one of the polarizations of gravitational waves.\footnote{Note that an Ostrogradsky mode associated with the higher derivatives of the scalar field shows up only in the even sector, and therefore it is irrelevant in the odd-mode analysis~\cite{Tomikawa:2021pca}. Similarly, in theories where the scalar field is nondynamical (e.g., the cuscuton model~\cite{Afshordi:2006ad} and its extension~\cite{Iyonaga:2018vnu,Iyonaga:2020bmm}, or a related theory of minimally modified gravity~\cite{DeFelice:2020eju,DeFelice:2022uxv}), the missing degree of freedom cannot be seen in the odd sector. As for the dipole odd-parity perturbations ($\ell=1$), they are nondynamical and related to the slow rotation of black holes~\cite{Mukohyama:2022skk}.}
In this case, we see that the condition~$h_2=0$ completely fixes the gauge degree of freedom, and hence this condition can be imposed at the level of Lagrangian~\cite{Motohashi:2016prk}.
Then, from Eq.~\eqref{eq:EFT_action}, one obtains a quadratic action for $h_0$ and $h_1$, which can be recast in terms of a single master variable~$\chi\coloneqq \dot{h}_1-\partial_\rho h_0$.
The master equation (i.e., the generalized RW equation) can be written in the form of a wave equation~\cite{Mukohyama:2023xyf},
	\begin{align}\label{eq:RW}
	\frac{\partial^2 \Psi}{\partial r_*^2}  - \frac{\partial^2 \Psi}{\partial \tilde{t}^2} -  V_{\rm eff}(r) \Psi = 0,
	\end{align}
where we have defined
    \begin{align}
    r_*\coloneqq \int \frac{{\rm d}r}{F}, \qquad
    \tilde{t}\coloneqq t+ \int \sqrt{\frac{1-A}{AB}}\frac{\alpha_T}{A+\alpha_T} {\rm d}r, \qquad
    \Psi\coloneqq \frac{r^3\chi}{\sqrt{1-A}\,(1+\alpha_T)^{3/4}},
    \end{align}
with
    \begin{align}
    F(r)\coloneqq \sqrt{\frac{B}{A}}\frac{A+\alpha_T}{\sqrt{1+\alpha_T}}, \qquad
    \alpha_T(r)\coloneqq -\frac{M_3^2(r)}{M_\star^2+M_3^2(r)}.
    \end{align}
The effective potential is given by
	\begin{align}\label{eq:RW_potential}
	V_{\rm eff}(r)
    =\sqrt{1+\alpha_T}\,F\left\{\sqrt{\frac{A}{B}}\frac{\ell(\ell+1)-2}{r^2}+\frac{r}{(1+\alpha_T)^{3/4}}\left[ F\left(\frac{(1+\alpha_T)^{1/4}}{r}\right)'\,\right]'\right\}.
	\end{align}
Here, the $r$-dependent parameter~$\alpha_T$ characterizes the relative difference between the squared sound speed of gravitational waves and that of light.
With $A=B$, the generalized RW equation presented here reduces to that in \S\ref{sec:RWeq}.
The condition for the absence of ghost/gradient instabilities can be written as~\cite{Mukohyama:2022skk}
    \begin{align}
    M_*^2>0, \qquad
    1+\alpha_T(r)>0.
    \end{align}
It should be noted that we have assumed $\alpha+M_3^2=0$ in deriving the above master equation, as otherwise the theory does not admit a slowly rotating black hole solution or the radial sound speed diverges at spatial infinity~\cite{Mukohyama:2022skk}.
Actually, this condition is automatically satisfied in shift-symmetric higher-order scalar-tensor theories~\cite{Takahashi:2019oxz,Tomikawa:2021pca,Mukohyama:2022skk}.

%%%%%%%%%%%%%%%%%%%%%%%%%%%%%%%%%%%%%%%%%%
\section{The explicit form of Eq.~(\ref{eqPDE})}\label{app:PDE_explicit_form}

Starting with the wave equation~\eqref{eq:RW_A=B}, we obtain the following PDE:
\begin{equation}
{\cal F}(r) {\cal A}(r) \partial_{\tilde{t}}^2 \Psi + {\cal B}(r) \partial_{\tilde{t}} \partial_{\bar{r}} \Psi + {\cal F}(r) {\cal C}(r) \partial_{\tilde{t}} \Psi + \partial_{\bar{r}}^2 \Psi + {\cal F}(r) {\cal V}(r) \Psi = 0,
\label{eqpdeapp}
\end{equation}
where
\begin{align}
&{\cal A}(r) \coloneqq -1 - \frac{\epsilon}{4} -\frac{r \epsilon}{12 \sigma}\,\text{sech}^2\left( \frac{r - \mu}{\sigma} \right) + \frac{\epsilon}{4} \tanh\left( \frac{r - \mu}{\sigma} \right),\\
&{\cal B}(r) \coloneqq \sqrt{\frac{r_0}{r}} \bigg[(2 + \epsilon)+ \frac{r \epsilon}{6 \sigma}\, \text{sech}^2\left( \frac{r - \mu}{\sigma} \right)\bigg] ,\\
&{\cal C}(r) \coloneqq - \frac{\sqrt{r_0}}{24 r^{3/2}} \left\{6 (2 + \epsilon)+ \frac{r \epsilon}{\sigma}\, \text{sech}^2\left( \frac{r - \mu}{\sigma} \right) \left[-1+\frac{4r}{\sigma}\tanh\left( \frac{r - \mu}{\sigma} \right)\right]\right\} ,\\
&{\cal V}(r) \coloneqq
\frac{1}{12 r^2} \left\{ \frac{r^2 (-r + r_0) \epsilon}{\sigma^3} \, \text{sech}^4\left( \frac{r - \mu}{\sigma} \right) \right. \nonumber \\
&\qquad\quad + 3 \bigg[ (-4 + \epsilon) \ell (1 + \ell) + \frac{3 r_0(4+3\epsilon)}{r} - \left( \ell (1 + \ell) - \frac{3 r_0}{r} \right) \epsilon \tanh\left( \frac{r - \mu}{\sigma} \right) \bigg] \nonumber \\
&\qquad\quad \left.+ \frac{r \epsilon}{\sigma} \, \text{sech}^2\left( \frac{r - \mu}{\sigma} \right) \left[ \left( (4 + \ell + \ell^2) - \frac{4 r_0}{r} \right) 
+ \frac{5 r - 2 r_0}{\sigma} \tanh\left( \frac{r - \mu}{\sigma} \right) + \frac{2 r (r - r_0)}{\sigma^2} \tanh^2\left( \frac{r - \mu}{\sigma} \right) \right] \right\}
,\\
&{\cal F}(r) \coloneqq -\frac{(r + r_0) \epsilon \, \text{sech}^2\left( \frac{r - \mu}{\sigma} \right)}{12 \sigma}
- \frac{r (-4 + \epsilon) + r_0 (4 + 3 \epsilon) + (-r + r_0) \epsilon \tanh\left( \frac{r - \mu}{\sigma} \right)}{4 r}
,
\end{align}
and we have defined the tortoise coordinate so that ${\rm d}\bar{r} = {\rm d}r/{\cal F}(r)$.
Note that the first term in Eq.~(\ref{eqpdeapp}) becomes zero at the effective horizon where ${\cal F}=0$. As such, the PDE does not satisfy the Courant-Friedrichs-Lewy (CFL) condition near the horizon.
To satisfy the condition, we perform the following coordinate transformation:
\begin{align}
{\rm d}\bar{r} \to {\rm d}x, \qquad
{\rm d} \tilde{t} \to {\rm d}\tilde{\tau} + \left( \frac{{\cal B}}{2} + \sqrt{\frac{{\cal B}^2}{4} + \frac{{\cal A} r_0}{r}} \right) {\rm d}x.
\end{align}
We then obtain the following PDE:
\begin{equation}
{\cal A}(r) \left( {\cal F} + \frac{r_0}{r} \right) \partial_{\tilde{\tau}}^2 \Psi + \partial_x^2 \Psi - \sqrt{\frac{{\cal B}(r)^2}{4} + \frac{{\cal A}(r) r_0}{r}} \partial_{\tilde{\tau}} \partial_x \Psi + [{\cal F}(r) {\cal C}(r) - \partial_x {\cal K}] \partial_x \Psi + {\cal F}(r) {\cal V}(r) \Psi =0,
\end{equation}
where ${\cal K} \coloneqq {\cal B}(r)/4 + \sqrt{{\cal B}^2(r)/4 + {\cal A}(r) r_0/r}$.
%%%%%%%%%%%%%%%%%%%%%%%%%%%%%%%%%%%%%%%%%%%%%%%%%%%%%%%%%%%%%%%%%%%%%%%%%%%%%%%%%%%%
%%%%%%%%%%%%%%%%%%%%%%%%%%%%%%%%%%%%%%%%%%%%%%%%%%%%%%%%%%%%%%%%%%%%%%%%%%%%%%%%%%%%

\bibliographystyle{mybibstyle}

\end{document}